\documentclass[%
 reprint,
superscriptaddress,
 amsmath,amssymb,
 aps,
floatfix,
]{revtex4-2}

\usepackage{graphicx}
\usepackage{dcolumn}
\usepackage{bm}
\usepackage{lipsum}
\usepackage{float}
\usepackage{hyperref}
\hypersetup{
   colorlinks=true,
   linkcolor=blue,
   filecolor=magenta,      
   urlcolor=cyan,
   pdftitle={Overleaf Example},
   pdfpagemode=FullScreen,
   }

\begin{document}

\preprint{APS/123-QED}

\title{Self-modulated multimode silicon cavity optomechanics}

\author{David Alonso-Tomás}
\email{david.alonso@ub.edu}
\affiliation{MIND-IN2UB, Departament d'Enginyeria Electrónica i Biomédica, Facultat de Física, Universitat de Barcelona, Martí i Franquès 1, Barcelona 08028, Spain}%
\author{Carlos Mas Arabí}
\affiliation{Institut Universitari de Matem\`{a}tica Pura i Aplicada, Universitat Polit\`{e}cnica de Val\`{e}ncia, 46022 (Val\`{e}ncia), Spain}%
\author{Carles Milián}
\affiliation{Institut Universitari de Matem\`{a}tica Pura i Aplicada, Universitat Polit\`{e}cnica de Val\`{e}ncia, 46022 (Val\`{e}ncia), Spain}%
\author{Néstor E. Capuj}
\affiliation{Depto. Física, Universidad de La Laguna, 38200 San Cristóbal de La Laguna, Spain}%
\affiliation{Instituto Universitario de Materiales y Nanotecnología, Universidad de La Laguna, 38071 Santa Cruz de Tenerife, Spain}%
\author{Alejandro Martínez}
\affiliation{Nanophotonics Technology Center, Universitat Politècnica de València, Camino de Vera s/n, 46022 Valencia, Spain}%
\author{Daniel Navarro-Urrios}
\email{dnavarro@ub.edu}
\affiliation{MIND-IN2UB, Departament d'Enginyeria Electrónica i Biomédica, Facultat de Física, Universitat de Barcelona, Martí i Franquès 1, Barcelona 08028, Spain}%

\date{\today}

\begin{abstract}
Multimode cavity optomechanics, where multiple mechanical degrees of freedom couple to optical cavity modes, provides a rich platform for exploring nonlinear dynamics and engineering complex interactions. In this work, we investigate the interplay between two mechanical modes with similar characteristics and a self-induced nonlinear modulation of intra-cavity power (self-pulsing) driven by free-carrier dispersion and thermo-optic effects in silicon. Notably, the self-pulsing dynamics adapts to the optomechanically induced perturbations from both mechanical modes, enabling simultaneous synchronous pumping and driving them into a stable state characterized by high-amplitude, self-sustained, and coherent oscillations. This result effectively overcomes the strong mode competition typically observed in modes with similar spatial distributions and frequency scales. Remarkably, this regime is achieved even when the mechanical frequencies do not satisfy a harmonic relation, leading to quasi-periodic or chaotic intra-cavity power dynamics, while the mechanical modes maintain coherent, high-amplitude oscillations. These results, supported by a numerical model that accurately predicts the system’s dynamics, open new pathways for the generation and control of multi-phonon coherent sources in chip-integrated silicon platforms.

\end{abstract}

\maketitle


\section{Introduction}
Cavity optomechanics \cite{Aspelmeyer}, which uses enhanced radiation pressure forces to couple optical and mechanical degrees of freedom, has become a powerful platform for exploring quantum physics and developing advanced technologies, such as ultrasensitive measurements and microwave-to-optical transduction \cite{Groblacher, Michael, sensing}. A key feature of this field is the ability to control optomechanical (OM) dynamics via radiation-pressure back-action, enabling phonon amplification under blue-detuned optical drives \cite{Kippemberg}. This has led to the development of coherent phonon sources ("phonon lasers") driven by DC optical fields, with applications in optical frequency comb generation \cite{Kuang, Mohammad}, sensing \cite{Liu}, and RF conversion \cite{Hossein, Mercade}.\\

While early studies focused on single-mode interactions, practical OM devices typically involve multiple optical and mechanical modes. This multimodal aspect has motivated efforts to explore systems with additional degrees of freedom, opening new avenues for studying nonlinear dynamics and engineering interactions \cite{Verhagen1, Verhagen2}. In multimodal OM platforms, parametric coupling between mechanical modes and a shared optical mode gives rise to effective interactions among the mechanical modes themselves. Under a strong blue-detuned optical drive, this can induce complex nonlinear phenomena such as synchronization \cite{Vitali, Wu}, chaos \cite{Djorwe}, and mode competition \cite{Zhang}. Similar to optical lasers, self-sustained oscillations in one mechanical mode can suppress gain in others, presenting a significant challenge for multimode phonon lasing (MML), particularly for mechanical modes with similar frequencies and spatial distributions \cite{Lawall}. \\

Several approaches have demonstrated the ability to mitigate this limitation and achieve MML. One strategy involves using mechanical modes with vastly different frequency scales, such as MHz and GHz, to ensure there is no mode competition between them \cite{Ryan, HuaDong}. Alternatively, for mechanical modes with similar frequencies, introducing external optical modulation at the intermodal frequency has shown promise. This modulation can facilitate Floquet phonon lasing in various OM platforms \cite{Laura, Guilhem} and regulate nonlinear phenomena such as bistability \cite{Guilhem2}. However, reliance on externally modulated optical drives adds complexity and cost to device design, potentially limiting scalability and hindering practical applications.\\

In this context, self-pulsing (SP), a self-induced modulation of intra-cavity power resulting from the interplay between free-carrier dispersion (FCD) and thermo-optic (TO) effects that can emerge in silicon OM cavities under low-power CW optical drives \cite{Johnson}, could play a key role in overcoming mode competition without external means. Unlike external modulation schemes, SP demonstrates remarkable versatility, capable of synchronizing with both externally driven \cite{Dani2} and optomechanically induced intracavity power modulations. Under the latter condition, it synchronously pumps the mechanical oscillator, driving it to a state of high-amplitude, self-sustained oscillations without the need for additional back-action amplification \cite{Dani}. Despite its promising capabilities, the potential of SP to manage multimode interactions and enable simultaneous mechanical amplification within a single OM device remains largely underexplored, with strong mode competition potentially limiting MML driven solely by SP dynamics \cite{HuaDong}. \\

Here, we address this issue by demonstrating and analyzing MML dynamics in a silicon OM nanobeam, operating without back-action amplification. Unlike previous studies, our research exclusively focuses on low-frequency, in-plane flexural modes of the nanobeam. These modes exhibit unique behavior, with mode competition directly influenced by intra-cavity power modulation induced by the self-pulsing (SP) mechanism.  In this context, we experimentally demonstrate a stable state in which two mechanical modes with similar spatial distributions and frequency scales coexist in phonon-lasing dynamics. To the best of our knowledge, this is the first demonstration of such coexistence arising intrinsically within the system, without the need for external modulation. In this regime, the SP mechanism adapts remarkably well to optomechanically induced perturbations caused by the high-amplitude, self-sustained oscillations of both mechanical modes. Notably, this adaptation persists even when the ratio between the mechanical frequencies deviates from exact integer values, enabling MML operation across periodic, quasi-periodic, and even chaotic regimes. These experimental findings are strongly supported by a theoretical macroscopic model, which accurately predicts the system’s dynamics and provides deeper insights into the underlying mechanisms.

\begin{figure*}[t]
    \centering    
    \includegraphics[width =\linewidth]{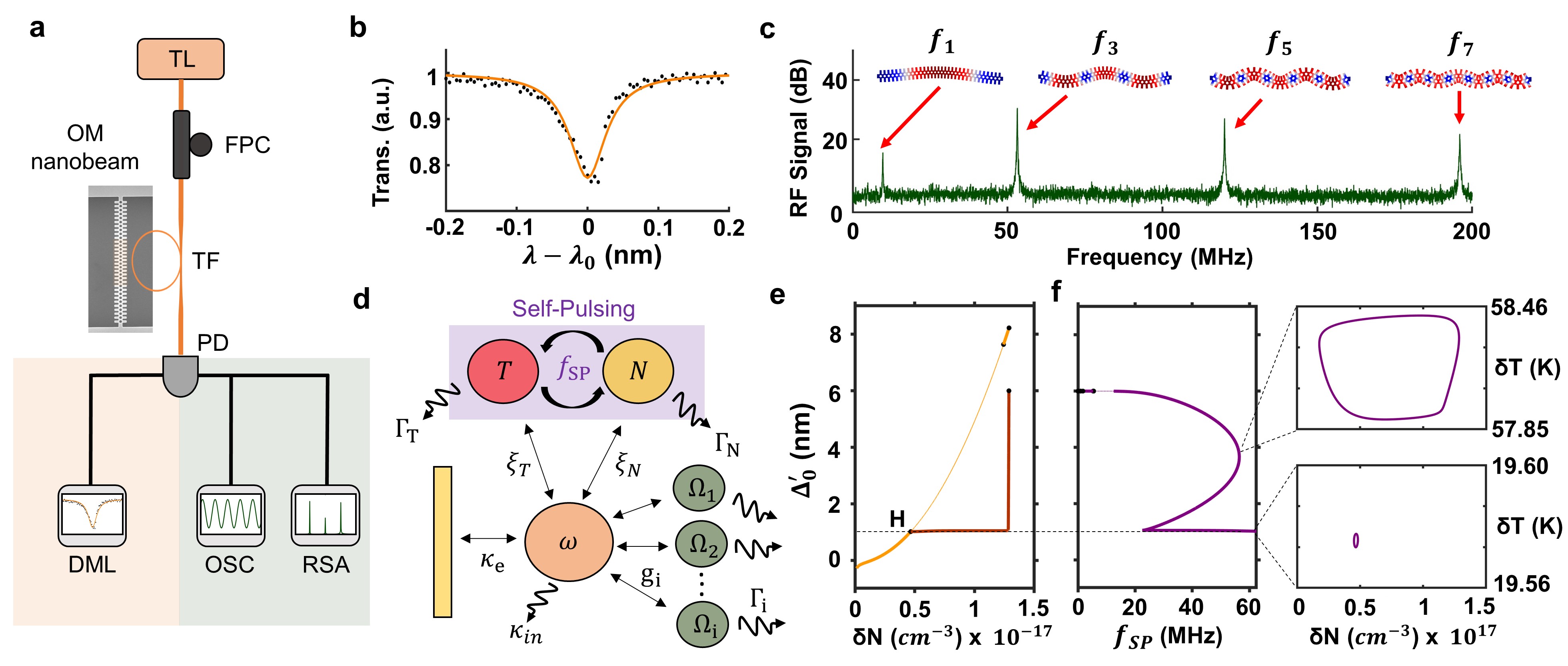}
    \caption{Characterization and description of the optomechanical system. \textbf{a} Experimental setup. Light from a tunable laser (TL) is coupled to an optomechanical nanobeam via a tapered fiber (TF), shaped into a microloop at its thinnest section. Polarization is optimized using a fiber polarization controller (FPC). The transmitted light is detected by a photodetector (PD) and analyzed using a digital multimeter (DML) and a radio-frequency spectrum analyzer (RSA). Temporal measurements are performed using a four-channel oscilloscope (OSC). \textbf{b} Normalized transmission recorded by the DML as the TL wavelength is tuned across the optical resonance at low input power ($P_{in}$ = 0.2 mW). \textbf{c} RF signal measured by the RSA after background subtraction. Several arrows link finite element method (FEM) simulations of the mechanical displacement for distinct in-plane odd antinode flexural modes to their corresponding RF peaks. \textbf{d} Schematic representation of key magnitudes and interactions in the system. The yellow bar symbolizes the optical fiber. \textbf{e} Steady-state solution (light orange) of $\delta N$ computed when sweeping the laser wavelength ($\lambda_L$) across the resonant wavelength ($\lambda_0$) from a blue-detuned position, i.e. $\Delta_0' = \lambda_L - \lambda_0 < 0$. In this case, no coupling with mechanical modes is considered. A black dashed line indicates the onset of a self-limit cycle at a Hopf bifurcation (H). Beyond this point, the maximum values of $\delta N$ in these periodic oscillations are shown by the brown thick line, with the light orange thin curve marking the instability region. Black dots indicates bifurcations. \textbf{f} Computed repetition rate of the self-pulsing limit cycle within the instability region. Insets display the cycle trajectory in the phase space ($\delta N$, $\delta T$) at its onset and at a later stage, after the cycle has fully developed.}
    \label{fig: setup}
\end{figure*}

\section{Results}
\subsection{Device and characterization}
The device under study is a silicon chip-integrated nanobeam optomechanical crystal cavity (see Appendix A for more details). In this structure, a confined optical mode is coupled to the mechanical degrees of freedom of the suspended nanobeam. Optical excitation is achieved by positioning the cavity in the near-field region of a tapered fiber \cite{Fabero}, which injects continuous wave (CW) laser light tuned to the resonant wavelength of the selected cavity mode (Fig. \ref{fig: setup}a). Our cavity exhibits an optical resonance at a wavelength of $\lambda_0 \sim 1519.4$ nm and a total optical decay rate of $\kappa/2\pi \sim 50$ GHz (see Fig. \ref{fig: setup}b), with a coupling efficiency of $\eta = \kappa_e / \kappa \sim 0.15$. Here, $\kappa_e$ represents the extrinsic optical decay rate, capturing external losses from the tapered fiber coupling channel. \\

The rapid dynamics in the output light, induced by radiation-pressure coupling with thermally activated mechanical modes \cite{Eichenfield}, can be detected and monitored using a spectrum analyzer. Figure \ref{fig: setup}c shows the RF signal of the detected light, where four in-plane mechanical modes with odd antinode numbers are observed in the frequency range between 0 and 200 MHz. The even modes are not observed, as they have a deformation minimum at the center of the structure, where the optical power is concentrated. The natural frequencies ($f_i$) of the mechanical modes at room temperature are $(\Omega_{1}, \Omega_{3}, \Omega_{5}, \Omega_{7})/2\pi = (9.71, 53.15, 120.05, 195.96)$ MHz with quality factors on the order of 500. This relatively low quality factor results from substantial air damping on the flexural modes under room pressure conditions.

\subsection{Theoretical framework}
The system studied in this work is schematically represented in Fig. \ref{fig: setup}d. Several mechanical modes of angular frequencies $\Omega_{i}$ and dissipation rates $\Gamma_{i}$ are coupled with vacuum optomechanical coupling rates $g_{i}$ to a single optical mode exhibiting an angular frequency $\omega$. The resonant frequency of the optical cavity can be altered by changes in temperature ($\delta T$) and free-carrier density ($\delta N$) through the TO and FCD effects, resulting in an effective detuning:
\begin{equation}
\Delta (\delta N,\delta T,\{x_i\}) = \Delta_0 + \xi_T \delta T + \xi_N \delta N + \sum_i G_i x_i
\end{equation}
where $\Delta_0 = \omega_0 - \omega_l$ represents the detuning of the frequency of the laser ($\omega_l$) relative to the cavity resonance at room temperature ($\omega_0$). Here, $\xi_{T} = \partial \omega/\partial \delta T$ and $\xi_{N} = \partial \omega/\partial \delta N$ are the linear coefficients associated to the TO and FCD effects. An additional term accounts for the contribution of the mechanical displacement of each mode $x_i$ on the detuning with a certain weight $G_i = \partial \omega/\partial x_i = g_i/x_{ZPF,i}$, where $x_{ZPF,i} = \sqrt{\hbar/2m_{eff,i}\Omega_i}$ is the mechanical zero-point fluctuation of mode i, and $m_{eff,i}$ is the effective mass of that mode. \\

In our case, the overall optical decay rate is significantly higher than the characteristic rates of the system. Under this condition, we can assume an instantaneous response of the intra-cavity photon number ($n$) to changes in $\delta N$, $\delta T$, and mechanical displacement of the set of mechanical modes $\{x_i\}$ and work with its steady-state solution derived from the input-output formalism \cite{Aspelmeyer}:

\begin{equation}
n (\delta N,\delta T,\{x_i\}) =  \frac{2\eta P_{in}}{\hbar \omega_l \kappa}\frac{1}{1 + 4\left(\frac{\Delta(\delta N, \delta T, \{x_i\})}{\kappa}\right)^2}
\end{equation}
Here, $P_{in}$ accounts for the incident optical power, while $\Delta$ is the implicitly time-dependent effective detuning defined in Eq. 1. \\ 

The macroscopic dynamics of $\delta N$ and $\delta T$ are governed by \cite{Dani2}:

\begin{equation}
\begin{aligned}
 \dot{\delta N} &= - \Gamma_{FC} \delta N + \alpha_{SPA}(N_0 - \delta N) \hspace{0.1cm} n(\delta N,\delta T,\{x_i\})\\
 &\dot{\delta T} = - \Gamma_{T} \delta T + \alpha_{FC} \delta N \hspace{0.1cm}  n(\delta N,\delta T,\{x_i\})\\
 \end{aligned}
\label{eq: system1}
\end{equation}
where the first equation describes the single-photon absorption (SPA) process involving $N_0$ intragap states per unit volume and a recombination time of 1/$\Gamma_{FC}$. The second equation accounts for linear photon absorption converted to heat via free-carrier absorption (FCA), with a thermal dissipation rate $\Gamma_T$. Here, $\alpha_{SPA}$ is the rate of increase in free-carrier density per photon per intragap state, while $\alpha_{FC}$ is the rate of temperature increase per photon per free-carrier density unit.\\

The dynamics of each mechanical mode can be described as a damped linear harmonic oscillator driven by radiation-pressure force exerted by intra-cavity photons:
\begin{equation}
\ddot{x}_j = -\Gamma_j\dot{x}_j - \Omega^2_j x_j + \frac{\hbar g_j}{x_{ZPF,j}m_j} \hspace{0.1cm} n(\delta N,\delta T,\{x_i\})
\end{equation}
where each equation for mode $j$ includes the term for intra-cavity photons, which depends on the entire set of mechanical modes. 
\\

The system described by equations (1-4) can be normalized and solved numerically for a fixed number of mechanical modes (see Appendix B). We first examine the case with zero mechanical modes, focusing on the dynamics of $\delta N$ and $\delta T$ alone. Figure \ref{fig: setup}e shows the steady-state value of $\delta N$ as $\Delta_0' = \lambda_L - \lambda_0$, expressed in terms of wavelength, is varied starting from a blue-detuned configuration.  \\

As the effective detuning decreases, photons enter the cavity, leading to increases in both $N$ and $T$. Here, the TO and FCD effects contribute to $\Delta$ in opposite directions, with the temperature effect dominating, shifting the optical resonance toward longer wavelengths. At a certain $\Delta_0'$ the system undergoes a Hopf bifurcation (H), where a self-sustained limit cycle, known as self-pulsing (SP), emerges between both magnitudes (see Fig. \ref{fig: setup}f). Despite the anharmonic nature of the cycle, it repeats periodically at a fundamental resonant frequency $f_{SP}$. Except for the two regions of cycle creation and dissipation, its frequency follows a parabolic dependence on $\Delta_0$', peaking in the middle region. \\

\begin{figure*}[t]
    \centering    
    \includegraphics[width =\linewidth]{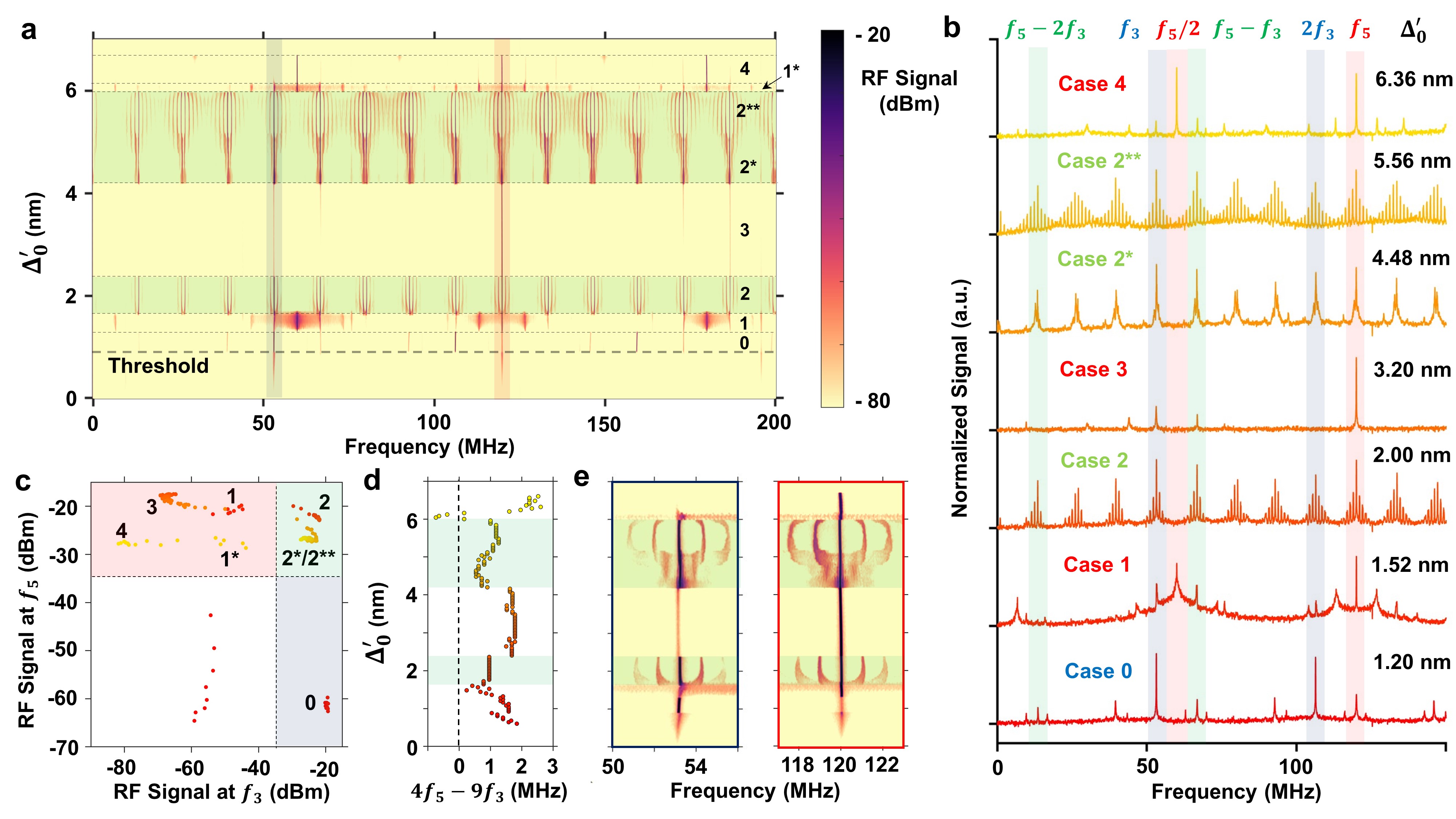}
    \caption{Experimental measurements of multimode lasing dynamics. \textbf{a} Contour plot of the  measured radio-frequency spectra as a function of the difference between the laser wavelength and the natural resonance position $\Delta_0' = \lambda_L - \lambda_0$. Dashed black lines indicate transitions between observed stable regimes, with regions of multimode lasing shaded in green. Each of the regions is denoted with a number from 0 to 4.\textbf{b} Selected RF spectra from the highlighted regions of interest, displayed on a normalized logarithmic scale, with the colormap transitioning from red to yellow as the wavelength detuning increases. The specific $\Delta_0'$ values corresponding to each case are also indicated. \textbf{c} Representation on a 2D map of the extracted RF signal corresponding to the tones at $f_3$ and $f_5$, with numbered markers indicating the highlighted regimes. \textbf{d} Plot of $f_d = 4f_5 - 9f_3$ derived from the monitored frequencies of M$_3$ and M$_5$, as a function of wavelength detuning. \textbf{e} Zoomed-in views around the natural mechanical frequencies of both modes, showing sidebands resulting from deviations from the integer mechanical frequency relation in both multimode regimes.}
    \label{fig: fig2}
\end{figure*}

From this point onward, we will focus on the interaction between the SP cycle and two mechanical modes, specifically those with frequencies f$_3$ and f$_5$ (denoted as M$_3$ and M$_5$), where mode competition for achieving lasing dynamics is observed.

\subsection{Multimode lasing experiment}
Considering the previously characterized geometry (see Fig. \ref{fig: setup}b,c), a similar sweep of the laser wavelength as mentioned in Fig. \ref{fig: setup}e is performed at an incident power of $P_{in} \sim 3$ mW. The result of this analysis is shown in Fig. \ref{fig: fig2}a, represented as a contour plot of the different measured RF spectra as a function of $\Delta_0'$. Several regions of interest beyond the threshold for mechanical lasing dynamics are highlighted and numbered from 0 to 4. For each region, a representative RF spectrum is extracted and plotted in Fig. \ref{fig: fig2}b. \\

\textit{Case 0 (Single-mode lasing of M$_3$)}: As the self-sustained limit cycle is created, it locks its repetition rate to $f_3$. Under these conditions, M$_3$ is synchronously pumped by the self-induced modulation of the intra-cavity power, entering single-mode lasing dynamics \cite{Dani}. Here, large excursion experienced by the optical resonance, driven by the combined effects of TO, FCD, and OM coupling, results in nonlinear signal transduction. This manifests as harmonics at integer multiples of the natural frequency of the mechanical mode (see \ref{fig: fig2}b). The RF intensity of the peaks at $f_3$ and $f_5$, which are directly related to the amplitude of the mechanical oscillations, is visualized in Figure \ref{fig: fig2}c. In particular, this regime is located in the blue area of the figure, associated with high-amplitude oscillations of M$_3$.\\

As the laser wavelength further increases, the frequency of the cycle adjusts, leading to transitions into various stable regimes spanning relatively wide regions of $\Delta_0'$. \\

\textit{Case 1 (Mode competition + Single-mode lasing of $M_5$)}: In this region, two stable solutions coexists, where M$_3$ attempts to sustain its lasing dynamics driven by the first harmonic of the SP cycle, while the mechanical mode at frequency $f_5$ (M$_5$) does the same with the second harmonic. The large difference between $f_5/2$ and $f_3$ results in a non-locked state, where the SP cycle is unable to drive both modes simultaneously, leading to phonon lasing of only mode M$_5$ (see 1 in Fig. \ref{fig: fig2}c) with a broadened response at $f_5/2$ resulting from the non-complete adaptation of the limit cycle.\\

\textit{Case 2 (MML)}: Counterintuitively, around $\Delta_0' = 1.7$ nm, high-amplitude self-sustained oscillations emerge in both mechanical modes, overcoming mode competition (see 2 in Fig. \ref{fig: fig2}b and \ref{fig: fig2}c). This regime originates from a stable state where both mechanical modes are simultaneously sustained by the modulation of intra-cavity power generated by the self-pulsing (SP) cycle. Notably, M$_5$ exhibits dominant oscillations, to which M$_3$ adapts by attempting to satisfy the condition $f_d = 4f_5 - 9f_3 = 0$ (see Figs. \ref{fig: fig2}d and \ref{fig: fig2}e). In practice, a frequency detuning of approximately 1 MHz remains. The nonlinear optomechanical transduction of this MML dynamics gives rise to a complex frequency comb, resulting from coherent beating between the unmatched RF signals. Sidebands at integer multiples of $f_d$ appear around the lasing peaks (see Fig. \ref{fig: fig2}e). These sidebands, also observed in other physical systems \cite{Steeneken2}, reflect quasi-periodicity stemming from the incommensurate relationship between the mechanical frequencies. The adaptation and simultaneous amplification of both mechanical modes suggest effective communication between them, mediated by the SP cycle. Further investigation into the origin of this MML stable state will be presented in subsequent sections.\\

\textit{Case 3 (Single mode lasing of M$_5$)}: The $f_{SP}$ curve reaches its maximum, enabling M$_5$ to enter the lasing regime driven by the fundamental harmonic over a broad range of wavelength detuning. Notably, the RF signal at $f_5$ increases compared to case 1 (see Fig. \ref{fig: fig2}c), as the mode is synchronously pumped at each oscillation rather than every two. Furthermore, the signal associated with M$_3$ decreases, indicating that in this regime, M$_5$ dominates, and the SP cycle is completely locked to it. \\

After reaching the maximum of its parabolic dependence, $f_{SP}$ decreases leading to a transition into a second region of MML (cases 2* and 2**), where the RF peaks associated with M$_3$ and M$_5$ exhibit reduced amplitudes (see 2*/2** in Fig. \ref{fig: fig2}c) in comparison to case 2. This decrease in phonon lasing amplitude suggests a less stable locked state of the SP cycle to both mechanical modes. This interpretation aligns with the significant drift in $f_d$ observed during this second MML regime (see Fig. \ref{fig: fig2}d), contrasting with the first one, where —despite not fully satisfying the condition $f_d = 0$— the frequency of the sideband remained stable. \\

Finally, after a transient regime (1*) similar to case 1, the SP cycle is able to completely lock to M$_5$ oscillation with its second harmonic, entering stable single-mode lasing (case 4). At this point, $f_{SP}$ no longer decreases sufficiently to revert to case 0, and the system destabilizes as the optical resonance shifts back toward its room-temperature position.

\begin{figure*}[t]
    \centering    
    \includegraphics[width =\linewidth]{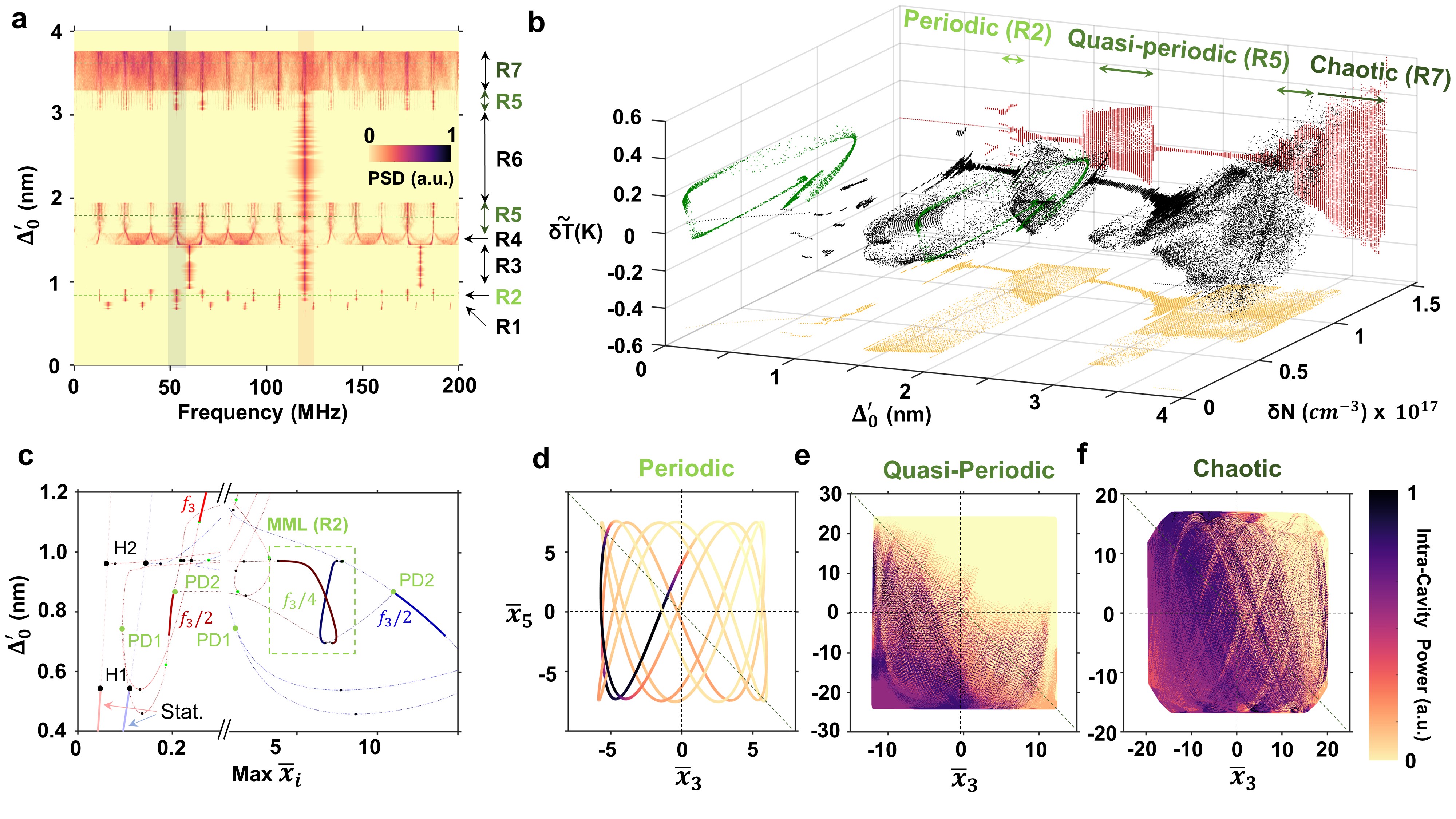}
    \caption{Numerical simulations of the self-pulsing system coupled to two mechanical modes. \textbf{a} Contour plot of the power spectral density (PSD), extracted from the fast Fourier transform (FFT) of temporal transmission traces, computed numerically by solving the model for different values of $\Delta_0'$. The sweep starts from a blue-detuned position. Key regimes are highlighted with double arrows and labeled "R" for differentiation from Fig. 2. RF peaks corresponding to M3 and M5 are highlight with red and blue areas, respectively. \textbf{b} Visualization of the system dynamics during the detuning sweep in (a), represented as a Poincaré map. The dynamics are depicted using $\delta N$ and $\tilde{\delta T} = \delta T - \text{mean}(\delta T)$, sampled at the minimum oscillation values of M$_3$, i.e., at the frequency $f_3$. A specific section at $\Delta_0' = 1.8$ is highlighted in green and projected into the ($\delta N, \tilde{\delta T}$) phase-space plane.  Projections for the entire sweep in the other phase-space planes are also shown for context. \textbf{c}  Steady-state solutions and periodic orbits emerging from a Hopf bifurcation (H2) depicted as functions of detuning, with black, prominent dots marking the bifurcations. The maximum value of the normalized amplitudes of the two mechanical modes ($\bar{x}_i = x_i/10^3 x_{zpf,i}$) are shown in blue (M$_3$) and red (M$_5$), respectively. Stable solutions are shown as thick continuous lines, unstable ones as thin dashed lines, and period-doubling bifurcations (PD) as green dots. Branches of different periods vary in saturation, with darker shades indicating lower repetition rates. A state characterized by high amplitudes in both mechanical modes, corresponding to the MML periodic regime, is enclosed within a green dashed box. \textbf{d-f} Phase-space plots of the normalized instantaneous amplitudes $\bar{x}_i$ for the two mechanical modes, illustrating the system's behavior in periodic (d), quasi-periodic (e), and chaotic (f) states. Each phase-space point is color-mapped to represent the intra-cavity power, extracted from $n$, at that specific location. A cubic interpolation is used for the non-periodic cases.}
    \label{fig: fig3}
\end{figure*}

\subsection{Numerical simulations}
In this section, we focus on comparing the dynamics observed in the experiments with those obtained from numerical simulations. The system of differential equations (1-4) is normalized and formulated as six first-order ordinary differential equations (ODEs) for the case with two mechanical modes. A sufficiently small timestep is chosen to accurately describe the system dynamics, along with a large timespan to ensure that a stable solution is reached. Temporal transmission traces can be generated from the solution for the intra-cavity photon number. Starting from a blue-detuned condition, a sweep is performed toward the cavity resonance, and the system is solved for each detuning value (see Appendix B for more details).\\

\subsubsection*{Detailed analysis}

Figures \ref{fig: fig3}a and \ref{fig: fig3}b present two different representations of the dynamics obtained from this analysis. In the first case, a fast Fourier transform (FFT) of the temporal transmission trace is computed for each detuning value, and the resulting RF spectra are represented as a contour plot. This representation allows for a direct comparison with Fig. \ref{fig: fig2}a. The second representation is a Poincaré map, sampling solutions for $\delta N$ and the deviation from the mean value of $\delta T$ at the different minima of M$_3$ oscillation.\\

Initially, before $\Delta_0'$ reaches 1 nm, two regimes emerge that are absent in Fig. \ref{fig: fig2}a. Regime R1 (see Fig. \ref{fig: fig3}a) involves the SP driving M$_3$ with its third harmonic, resulting in RF peaks at integer multiples of $f_3/3$. These fractional states are commonly observed in similar studies of these cavities and are well described by the model \cite{Dani}. Subsequently, a periodic solution with a repetition rate of $f_3/4$ emerges in regime R2, where both mechanical modes exhibit high-amplitude, coherent oscillations, suggesting a periodic MML regime. To further investigate the emergence of these periodic solutions, we perform a bifurcation analysis of the system’s differential equations (see Appendix B). The steady-state mechanical amplitudes of both modes as a function of detuning are shown in Fig. \ref{fig: fig3}c. Two Hopf bifurcations (H1 and H2) are identified, marking the onset of lasing for each mechanical mode. Periodic orbits emerging from H2 reveal a variety of bifurcations, including a period-doubling bifurcation (PD1) that results in a stable state for $0.72 < \Delta_0' < 0.87$ with a repetition rate $f_3/2$.  A subsequent period-doubling bifurcation (PD2) leads to a stable state with repetition rate $f_3/4$, where both mechanical modes exhibit comparable high amplitudes ($x_i \sim 7 \cdot 10^3 x_{\text{zpf},i}$). This confirms that a periodic MML state is a stable solution of the system. Notably, both cases (R1 and R2) are visualized as three or four distinct points, respectively, in the phase space ($\delta N$, $\tilde{\delta T}$) of the Poincaré map for each detuning value (see Fig. \ref{fig: fig3}b). \\

Following these initial regimes, single-mode phonon lasing of M$_5$ is achieved (R3), with the SP cycle locked at half its frequency. Two solutions coexist in this detuning region: M$_5$ driven by the second harmonic of SP and M$_3$ driven by the first harmonic (see Appendix B). The preference between them may vary depending on specific parameters or the position in phase space. Indeed, the main experiment shows phonon lasing for M$_3$ instead (case 0). Subsequently, the system enters a transient state (R4), similar to case 1, where a cycle begins to form in the Poincaré map (see Fig. \ref{fig: fig3}b). This transition leads to an MML state (R5) in which both mechanical modes exhibit high-amplitude oscillations. This state is analogous to the one observed in the main experiment (case 2), where the frequency ratios between the two mechanical modes do not perfectly match, i.e., $4f_5 \neq 9f_3$. As discussed earlier, the frequency analysis suggests quasi-periodicity, which is corroborated by the cycle observed in the Poincaré map. Interestingly, this also highlights the extreme adaptability of SP, which, mediated by complex nonlinear interactions through intra-cavity photons, adjusts its cycle to synchronize with the mechanical perturbation, even if this requires unfolding or progressing at varying rates within each period (see supplementary video 1). \\

Following this first MML quasi-periodic regime, the SP drives mode M$_5$ into the lasing regime via its first harmonic (R6), analogous to case 3 in Fig. \ref{fig: fig2}b. Here, $f_{SP}$ reaches the peak of its parabolic dependence and then begins to decrease, ultimately recovering the same MML quasi-periodic state observed at lower $\Delta_0'$ values (R5). Interestingly, before destabilization, the system dynamics exhibit a large frequency broadening (R7) while maintaining the high-amplitude, coherent oscillations of the mechanical modes. This chaotic behavior in the SP cycle \cite{chaos} is evident in the Poincaré map (see Fig. \ref{fig: fig3}b), where scattered points, rather than a defined cycle, appear in the phase space ($\delta{N}, \tilde{\delta T}$). Experimental evidence of this MML regime is provided in Appendix C. \\

\subsubsection*{Origin of the MML states}

To better understand the pumping mechanism of both mechanical modes under these complex regimes, we analyze the intra-cavity photon power as a function of the instantaneous amplitudes of the oscillators. Figures \ref{fig: fig3}d-f illustrate this analysis, where the Lissajous curves of both mechanical oscillators are plotted over a temporal trace spanning several microseconds. A colormap is applied to each point, representing the intra-cavity power at that specific time. Interestingly, a clear asymmetry in the intra-cavity power distribution is observed for all MML states, with higher power concentrated in the bottom-left region of the maps. This symmetry breaking result in a net driving force that sustains and amplifies the oscillations, ultimately leading to a self-sustained, coherent regime. Notably, this behavior is not restricted to periodic solutions. The adaptability of the SP enables a quasi-synchronous pumping of both mechanical oscillations even under quasi-periodic or chaotic regimes (see Appendix B and Supplementary Video 2 for further clarification). 

\subsection{Periodic and Quasi-periodic solutions}
\begin{figure*}[t]
\centering    
\includegraphics[width= \linewidth]{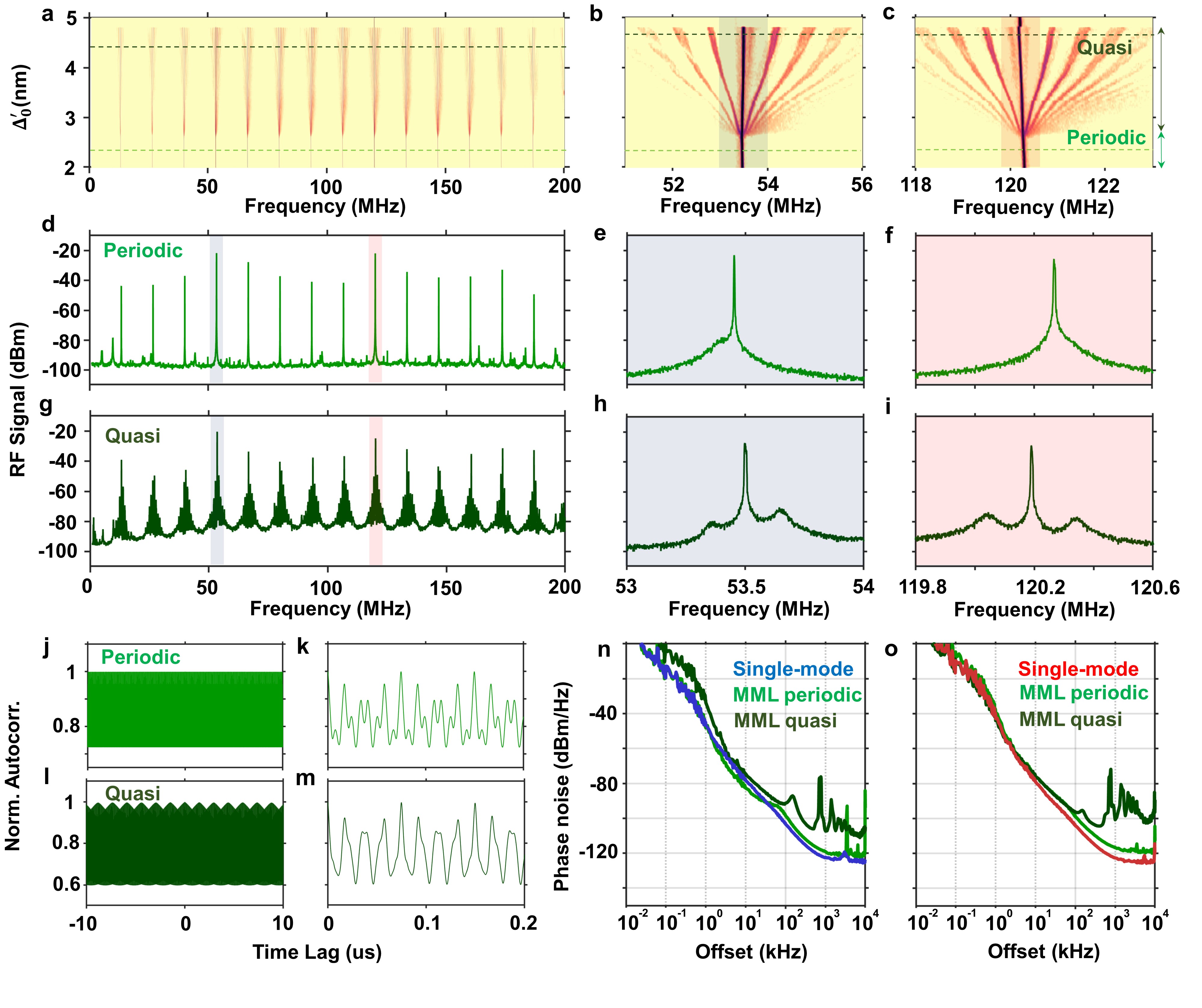}
\caption{Experimental measurements of an OM nanobeam exhibiting periodic and quasi-periodic MML. \textbf{a} Contour plot of the RF spectra as a function of the detuning between the laser wavelength and the natural resonance position, $\Delta_0'$. Light and dark green dashed lines highlight the cases analyzed in subsequent panels. \textbf{b-c} Magnified views around $f_3$ and $f_5$ throughout the detuning excursion in (a). \textbf{d} RF spectrum corresponding to the periodic state highlighted in (a). A semitransparent area marks the RF tones associated to both mechanical modes: blue for M$_3$ and red for M$_5$. \textbf{e-f} Magnifications around $f_3$ and $f_5$ for the specific case presented in (d). \textbf{g} RF spectrum corresponding to the quasi-periodic state highlighted in (a). \textbf{h-i} Magnifications around $f_3$ and $f_5$ for the specific case presented in (g). \textbf{j-m} Computed autocorrelation from temporal traces measured with the oscilloscope in each MML regime.  Panels (k) and (m) are magnified views over a small time lag of (j) and (l), respectively. \textbf{n-o} Phase noise of the RF tones corresponding to M$_3$ and M$_5$ in single-mode operation (red), periodic MML (light green), and quasi-periodic MML (dark green). \label{fig: phasenoise}}
\end{figure*}

Figure \ref{fig: fig2} presents experimental measurements demonstrating the coexistence of high-amplitude, self-sustained oscillations in two flexural modes of a silicon nanobeam. The incommensurate relationship between the mechanical frequencies results in an optomechanically induced quasi-periodic modulation of the intra-cavity power, to which the SP cycle adapts. Numerical simulations align with this observation but also suggest that a periodic MML state is achievable when the frequencies of the mechanical modes lock to a specific ratio. Here, we present an experimental case where $f_3$ and $f_5$ are sufficiently close to the condition $f_d = 0$ to lock into a periodic MML regime. A similar experiment to that shown in Figure \ref{fig: fig2}a is depicted in Fig. \ref{fig: phasenoise}a, focusing on a region where a periodic MML regime transitions into quasi-periodic oscillations. This transition is clearly observed in the magnified views around $f_3$ and $f_5$ (see Figs. \ref{fig: phasenoise}b and c, respectively), with the emergence of sidebands around the main RF tones as $\Delta_0'$ increases.\\

A representative case is selected for each regime, and several properties are analyzed. First, the RF spectrum of the periodic MML regime is presented in Fig. \ref{fig: phasenoise}d, showing a clear comb with a repetition rate of $f_3/4$, resulting from the beating between the RF signals associated with the non-linear optomechanical transduction of the high-amplitude, self-sustained oscillations of the mechanical modes. Magnified views around the RF peaks are depicted in Figs. \ref{fig: phasenoise}e and \ref{fig: phasenoise}f. Similar to the experimental measurement presented in Fig. \ref{fig: fig2}, both oscillators adapt to satisfy the condition $f_d = 0$, with $M_3$ exhibiting a greater extent of adaptation than $M_5$. In this case, the locked state is achieved, where a high-amplitude RF peak is observed in both frequency ranges ($f_3 = 53.45$ MHz and $f_5 = 120.26$ MHz), superimposed on a broad signal that appears at the natural oscillation frequencies, with a bandwidth comparable to the dissipation rate of the mechanical modes ($\Gamma \sim 0.2$ MHz). This resulting spectrum can be understood in terms of the gain mechanism. The synchronous pumping mediated by SP injects energy at frequencies related harmonically (9:4). Here, the mechanical density of states at the resonant frequencies amplifies the response of the modes, resulting in simultaneous phonon lasing. \\

Fig. \ref{fig: phasenoise}g presents the RF spectrum extracted from the data set corresponding to Fig. \ref{fig: phasenoise}a (see the dark green dashed line) after transitioning into the MML quasi-periodic regime. Magnified views around the RF tones at the mechanical mode frequencies (Figs. \ref{fig: phasenoise}h and \ref{fig: phasenoise}i) reveal similar behavior. High-amplitude, sharp peaks are observed in both frequency regions; however, in this case, they do not satisfy the harmonic relationship ($f_3 = 53.50$ MHz and $f_5 = 120.19$ MHz). Interestingly, sidebands appear at 145 KHz from both peaks, in addition to those observed at integer multiples of $f_d$.  \\

\subsubsection*{Autocorrelation}

Subsequently, temporal traces with a resolution of 50 ps and spanning 100 $\mu$s are measured with the oscilloscope for each of the selected states.  Figures \ref{fig: phasenoise}j–m show different time windows of the computed autocorrelation from the measured traces. Both the periodic and quasi-periodic MML regimes maintain a high autocorrelation value across the entire time lag. Interestingly, this magnitude exhibits a pattern resembling the beating between two coherent oscillations with the relation 9/4 (see Fig. \ref{fig: phasenoise}k), further confirming the MML regime and suggesting phase locking between the two mechanical modes. On the other hand, periodic oscillations on a larger timescale are clearly observed in the quasi-periodic MML regime (see Fig. \ref{fig: phasenoise}l), while the interference pattern is likely disrupted due to the unmatched frequency ratio between the mechanical modes (Fig. \ref{fig: phasenoise}m).  \\

\subsubsection*{Phase-noise}

Finally, high-precision linewidth measurements are obtained using a phase noise analyzer (see Figs. \ref{fig: phasenoise}n and \ref{fig: phasenoise}o). Oscillators operating in the MML periodic regime (light green) exhibit phase noise comparable to that observed during single-mode operation (red and blue curves). In contrast, the quasi-periodic regime shows degraded phase noise at high-frequency offsets compared to the periodic case. This degradation is associated with the imperfect adaptation of the SP cycle to the complex beating caused by unmatched RF tones. Notably, the RF peak corresponding to the mode exhibiting greater adaptation (M3) in this regime also shows higher frequency noise at low offsets compared to the periodic case. This increased noise is primarily attributed to RF tone jitter.

\subsection{Probing the mechanical signal}

\begin{figure}[t]
    \centering    
    \includegraphics[width =\linewidth]{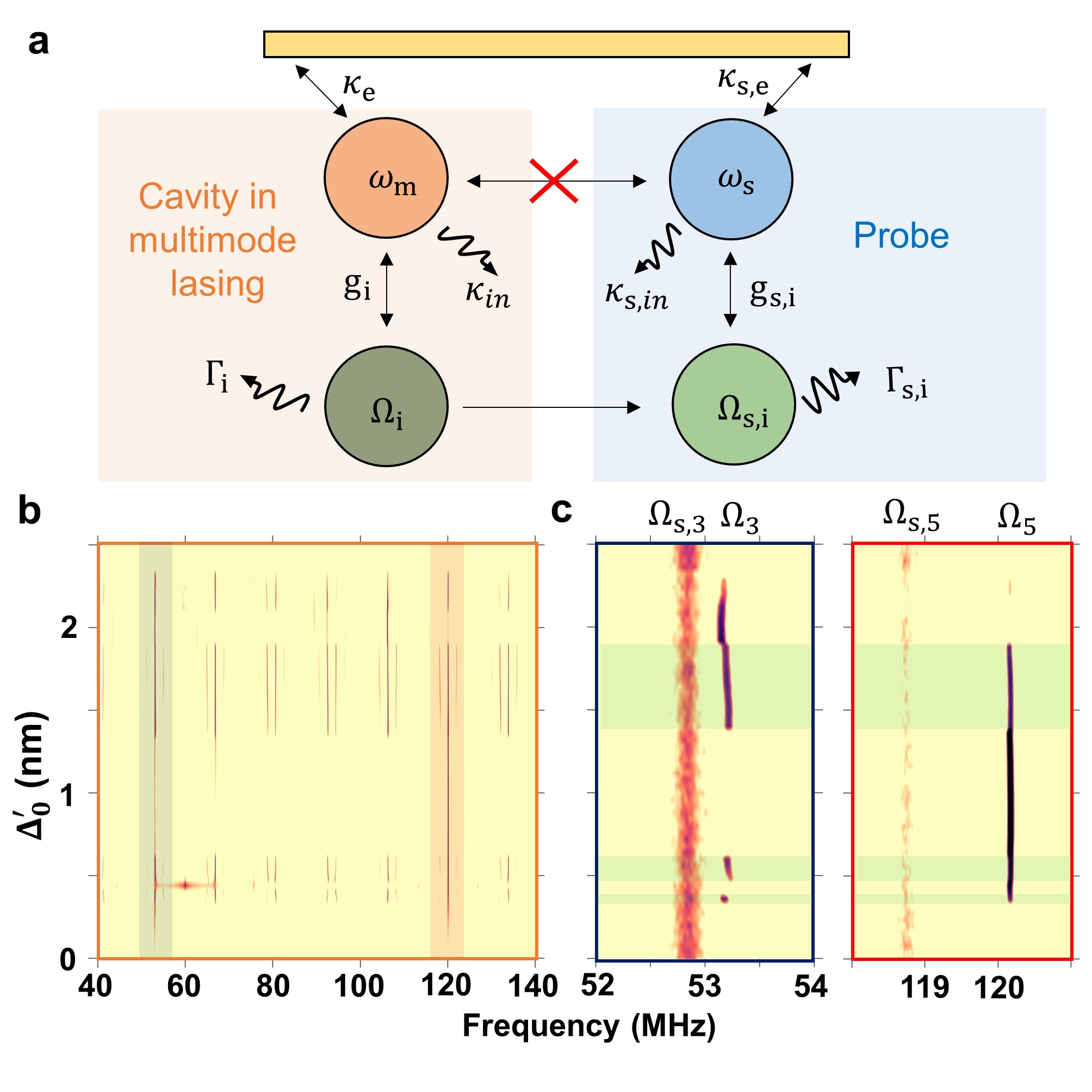}
    \caption{Experiment for probing the mechanical signal of the cavity in multimode phonon lasing dynamics.\textbf{a} Schematic representation of the mechanical probing setup. Both optical modes are isolated, while the mechanical modes interact through a common frame.  \textbf{b} Contour plot of the measured RF spectra for the optical channel corresponding to the cavity that supports the MML regime. Signals associated with the mechanical frequencies of M$_3$ and M$_5$ are highlighted in blue and red, respectively. \textbf{c} Similar representation as in (b), for the optical channel in resonance with the probe cavity. Two magnified graphs are presented, focusing on frequencies $f_3$ and $f_5$. }
    \label{fig: fig4}
\end{figure}

Experimental measurements from previous sections have revealed the emergence of an optical RF comb, where tones associated with the mechanical motion of both oscillators exhibit increased amplitude and reduced linewidth. This suggests that both mechanical modes achieve high-amplitude, self-sustained coherent oscillations simultaneously. Two key observations support this conclusion:
\begin{itemize}
    \item The computed autocorrelation of the temporal trace in the MML regime displays a beating pattern between two oscillators with a frequency ratio of 9/4.
    \item The numerical model presented demonstrates dynamics in optical transmission that closely resemble those observed in the experiment, while also predicting high-amplitude, self-sustained oscillations in the mechanical modes.
\end{itemize}  

Here we provide a technique to experimentally access to the mechanical signal, further confirming the MML regime. The key concept consist on using a second OM cavity as a mechanical probe (see sketch in Fig. \ref{fig: fig4}a). This configuration has been explored in previous works \cite{Arregui}, where the second cavity is positioned 2 $\mu$m away, enabling optical excitation of both cavities by placing a tapered fiber in between. The cavities support separate optical resonances, ensuring no optical crosstalk. The common anchoring on the frame allows the probe cavity to sense the dynamics of the main structure, which exhibits large deformation amplitudes in the lasing regime.\\

Two tunable lasers are tuned in resonance with each respective optical mode. The probe laser wavelength is fixed at a detuning where no self-pulsed dynamics are activated in the secondary cavity, allowing only the transduction of its thermally activated mechanical modes. Subsequently, the main laser wavelength is swept from lower to higher values around its resonance position, enabling the emergence of mechanical lasing dynamics. For each wavelength detuning value, two measurements are taken: one from the main channel (Fig. \ref{fig: fig4}b) and the other from the probe channel (Fig. \ref{fig: fig4}c). This is achieved by using a software-controlled wavelength filter before the photodetector. As in previous experiments and simulations, two regions of MML are observed, both before and after the maximum of $f_{SP}$ is reached. Fig. \ref{fig: fig4}c shows magnified views around the frequencies of both mechanical modes observed in the probe channel. In each panel, two main signals can be distinguished. The broader, lower-intensity signal corresponds to the transduction of the natural mechanical mode of the probe cavity. To its right, a narrower, higher-intensity peak represents the mechanical perturbation generated by the main cavity. The detuning range over which these peaks appear aligns with the dynamics observed in the main channel (Fig. \ref{fig: fig4}b), indicating high-amplitude, self-sustained oscillations of both mechanical modes in the MML regime. \\

\section{Discussion}
In this work, we have theoretically and experimentally studied an optically self-modulated multimode optomechanical platform. In this system, two mechanical modes interact with a self-sustained limit cycle generated by the interplay of free-carrier dispersion (FCD) and thermo-optic (TO) effects via intra-cavity photons. Simultaneously, the self-induced modulation produced by this tunable cycle influences the dynamics of the mechanical oscillators through radiation pressure forces. \\

The mutual interaction between these oscillators results in a stable state, where the self-induced nonlinear modulation of the radiation pressure force drives both mechanical resonators into simultaneous, self-sustained, and coherent oscillations. Remarkably, this occurs even though both resonators are in-plane flexural modes with resonant frequencies in the same order of magnitude and similar spatial distributions, effectively overcoming the challenge of mode competition \cite{Lawall} without requiring external means. Experimental observations reveal a dominant oscillator to which a follower adapts, approaching or satisfying a frequency ratio of 9:4 between their mechanical modes. Under the latter condition, the RF tones generated by the optomechanically transduced lasing dynamics produce a beating pattern that facilitates effective communication between the modes, locking in phase. However, when this ratio do not satisfy an integer relationship, simultaneous amplification of both modes becomes less evident. \\

In this context, SP plays several critical roles. First, it provides an optical resonance excursion large enough to sustain significant intracavity power variations, which the two mechanical modes in our system cannot independently achieve in the simultaneous lasing regime. Second, SP adapts to the perturbations generated by the optomechanically transduced oscillations, enabling a synchronous pumping that result in amplification of both mechanical oscillators even when the relation between the mechanical modes is not fully satisfied. \\

Together, these results establish an excellent platform for realizing tunable multi-phonon coherent sources, whose internal dynamics are mediated by the system itself. This approach holds significant potential for compact and adaptive devices capable of generating frequency combs at low repetition rates. Additionally, the observed asymmetry between intracavity power and mechanical amplitudes in the MML state opens intriguing avenues for exploring novel regimes in which multiple mechanical modes are simultaneously amplified, providing insights into the interplay between symmetry breaking and multimodal dynamics in optomechanical systems. While this work focuses on mode competition between two mechanical modes, experimental observations (see Appendix D) suggest that simultaneous phonon lasing involving more than two mechanical modes is possible when their frequency ratios approximate integer relationships. Future work will aim to introduce nonlinearity into the mechanical oscillators \cite{Steeneken}, which could further enhance the adaptability required for MML operation in periodic regimes.


\section*{Acknowledgments}
This work was supported by the MICINN projects ALLEGRO (Grants No. PID2021-124618NB-C22 and  PID2021-124618NB-C21) and MOCCASIN-2D (Grant No. TED2021-132040B-C21).

\section*{Appendix}
\subsection{Optomechanical nanobeams}
The structures investigated in this work are one-dimensional silicon optomechanical (OM) crystal cavities, which are free-standing nanobeams designed to function simultaneously as photonic and phononic crystals. A localized defect region within these crystals is created through adiabatic modulation of the unit cell parameters, enabling the confinement of optical and mechanical modes. Specifically, the crystal geometry is engineered to act as a complete mirror for optical frequencies around 200 THz \cite{Jordi}. The devices were fabricated using standard silicon-on-insulator (SOI) wafers provided by SOITEC. The  se wafers consisted of a 220 nm thick silicon layer with a resistivity ranging from 1 to 10 $\Omega$ cm$^{-1}$ and a p-type doping concentration of approximately $10^{15}$ cm$^{-3}$. A 2 $\mu$m thick buried oxide layer was present beneath the silicon layer. The design was patterned onto a 100 nm thick poly-methyl-methacrylate (PMMA) resist using electron beam lithography. This pattern was subsequently transferred into the silicon layer via Reactive Ion Etching (RIE). Finally, buffered hydrofluoric acid (BHF) was used to etch away the buried oxide layer, releasing the fabricated nanobeam structures.
\subsection{Numerical simulations of free-carrier and temperature dynamics in a multimode optomechanical system}

\subsubsection*{Normalization}
The system of differential equations presented in Eq. \ref{eq: system1} exhibits multiple distinct numerical scales due to the interplay between the free-carrier density and temperature magnitudes. To analyze this system effectively, we normalize it using several characteristic magnitudes of the system: 

\begin{center}
\begin{tabular}{ c c }
 $\bar{\tau}$ & $\tau \alpha$ \\
 $\bar{\Gamma}_{FC}$ & $\Gamma_{FC}/\alpha$ \\
 $\bar{\Gamma}_T$ & $\Gamma_T/ \alpha$ \\
 $\bar{N}$ & $S N/N_0$ \\  
 $\bar{T}$ & $T/r$ \\
 $\bar{\xi_T}$ & $\xi_T(r/\kappa)$ \\
 $\bar{\xi_N}$ & $\xi_N(N_0/\kappa)$\\
 $\bar{\Delta_0}$ & $\Delta_0/\kappa$ \\
 $\bar{g}$ & $g/\alpha$ \\
 $\bar{\Omega}$ & $\Omega/\alpha$ \\  
 $\bar{x}$ & $x/(S x_{ZPF})$ \\

\end{tabular}
\end{center}

Here, $\alpha = \alpha_{SPA} n_c$, where $n_c = 2 \cdot 10^5$ represents a specific intra-cavity photon normalization constant. Increments in the free-carrier density and temperature, $\delta N = N$ and $\delta T = T$, are normalized by the density of intra-gap states ($N_0$) and $r = \alpha_{FC} N_0 / \alpha_{SPA} S$, respectively. The mechanical displacement ($x$) is normalized by the mechanical zero-point fluctuations ($x_{ZPF}$) and $S = 10^3$. With this normalization, the system of equations (3–4) can be expressed as:

\begin{figure*}[t]
\centering    
\includegraphics[width= \linewidth]{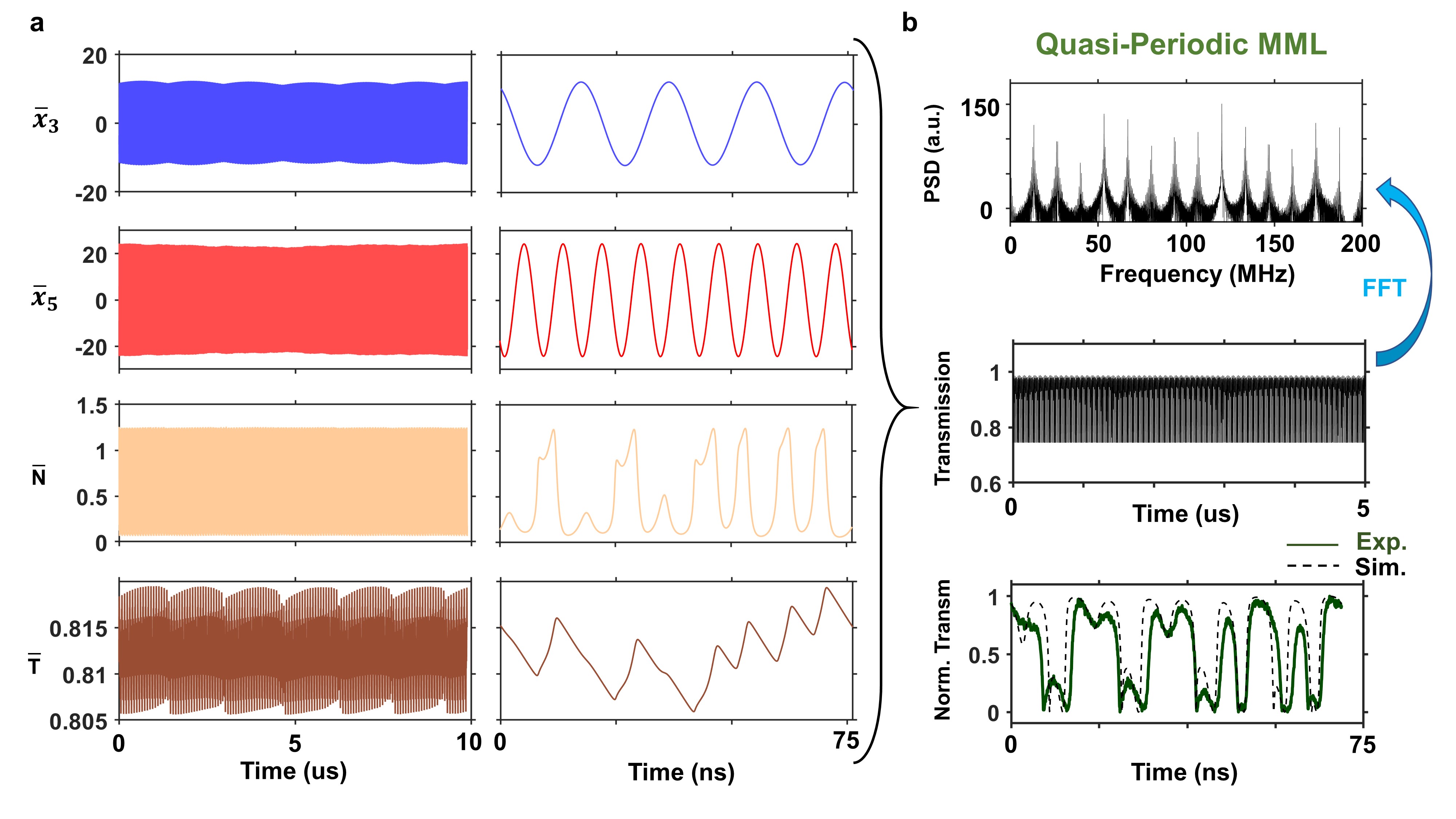}
\caption{\textbf{a} Computed temporal traces of several state variables of the model exhibiting the quasi-periodic multimode phonon lasing regime. Each row shows a long-timespan trace alongside a magnified view during an SP quasi-cycle. \textbf{b} Computed transmission temporal trace derived from the state variables over a span of 5 $\mu$s. A comparison of a single quasi-cycle with an experimental temporal trace measured via oscilloscope is shown below. The corresponding RF spectrum, obtained through the fast Fourier Transform of the trace, is also presented. \label{fig: quasi}}
\end{figure*}

\begin{subequations}
\begin{align}
& \frac{d \bar{N}}{d \bar{\tau}} = - \bar{\Gamma}_{FC} \bar{N} + S \cdot n(\bar{N},\bar{T},\{\bar{x}_i\})\\
& \frac{d \bar{T}}{d \bar{\tau}} = - \bar{\Gamma}_{T} \bar{T} + \bar{N} \cdot n(\bar{N},\bar{T},\{\bar{x}_i\})\\
& \frac{d \bar{x}_i}{d \bar{\tau}} = \bar{y}_i\\
& \frac{d \bar{y}_i}{d \bar{\tau}} = -  \frac{\bar{\Omega}_i}{Q} \bar{y}_i - \bar{\Omega}_i^2 \bar{x}_i + \frac{2 \bar{\Omega}_i \bar{g}_i n_c}{S} n(\bar{N},\bar{T},\{\bar{x}_i\})
\end{align}
\label{eq: system2}
\end{subequations}

where $N_0$ is set to be much higher than $N$, ensuring it acts as a source for the excitation of free carriers to the conduction band via single-photon absorption, which has been shown to be the dominant mechanism in our devices (see \cite{Dani2}). Here, the normalized intra-cavity photon number $n(\bar{N}, \bar{T},\{\bar{x}_i\})$ is given by:

\begin{equation}
n(\bar{N},\bar{T},\{\bar{x}_i\}) = \frac{\eta P}{1 + 4\bar{\Delta}^2} \hspace{0.5cm}
\end{equation}\\
where $\bar{\Delta} = \bar{\Delta} (\bar{N},\bar{T},{\bar{x}_i})$ is the normalized effective detuning:
\begin{equation}
\bar{\Delta}(\bar{N},\bar{T},\{\bar{x}_i\}) = \bar{\Delta}_0 + \bar{\xi}_T \bar{T} + \bar{\xi}_N  \bar{N} + \frac{S\alpha}{\kappa} \left(\sum_i \bar{g}_i \bar{x}_i \right)
\end{equation}

and $P = 2P_{in}/(\hbar \omega_L \kappa n_c)$ the normalized input power. \\

An expression for the transmission coefficient can be derived from the input-output formalism:

\begin{equation}
\Pi(\bar{N},\bar{T},\{\bar{x}_i\}) \approx 1 -  \frac{2\eta (1 - \eta)}{1 + 4\bar{\Delta}^2} = 1 - \frac{2(1- \eta)}{P} n(\bar{N},\bar{T},\{\bar{x}_i\}).
\label{eq: transm}
\end{equation}
The parameters of the model are extracted from experimental measurements \cite{Johnson,properties1, properties2,properties3,properties4} and simulations. Tables 1 and 2 presents the values used for the numerical simulations presented in the work. \\

\begin{figure*}[t]
\centering    
\includegraphics[width= \linewidth]{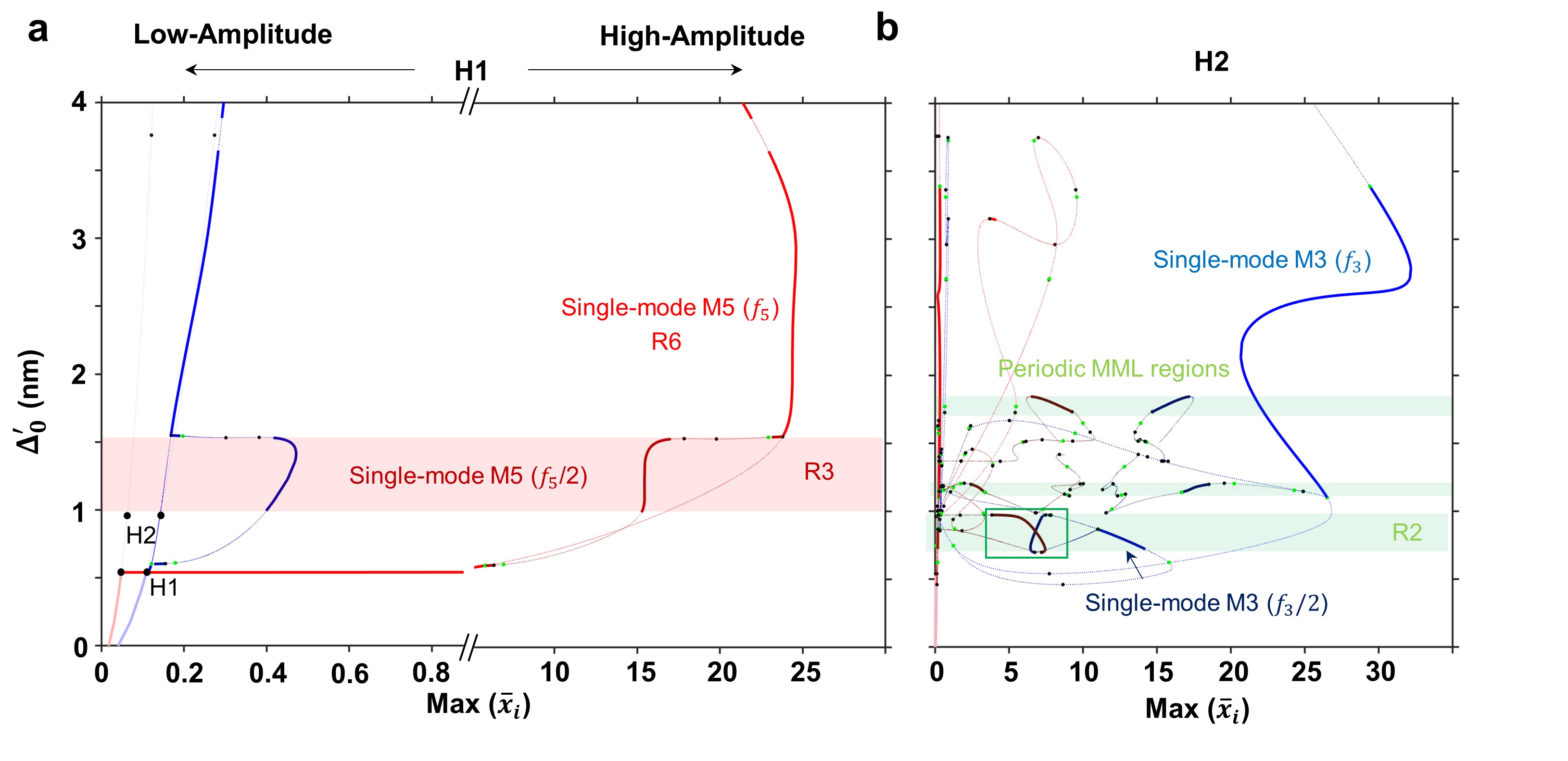}
\caption{Stationary and periodic orbits emerging from two Hopf bifurcations as a function of $\Delta_0'$. The maximum value of the normalized instantaneous amplitudes of the two mechanical modes are shown in blue (M$_3$) and red (M$_5$), respectively. Stable solutions are represented by thick continuous lines, while unstable ones are depicted as thin dashed lines. Period-doubling bifurcations (PD) are indicated by green dots. Branches with different periods are distinguished by varying saturation levels, with darker shades representing lower repetition rates. \textbf{a} Case corresponding to H1, associated with the lasing onset of M5. The range of detuning where the solution with a repetition rate of $f_5/2$ exists is highlighted with a red area. \textbf{b} Case corresponding to H2, associated with the lasing onset of M3. Several stable solutions where both modes coexist with high-amplitude are highlighted by a green area, corresponding to a repetition rate of $f_3/4$. \label{fig: bifurcations}}
\end{figure*}

\begin{table}[htb]
\centering
\caption{Parameters of the model before normalization.}
\begin{ruledtabular}
\begin{tabular}{lccc} 
Parameter & Value & Parameter & Value \\ \hline
$\alpha_{SPA}$ (s$^{-1}$)  & 40  & $\alpha_{FC}$ (K m$^{3}$ s$^{-1}$) & 1.65 $\cdot$ 10$^{-14}$\\
$N_0$ (m$^{-3}$) & 10$^{20}$ & $\kappa$/2$\pi$ (GHz) & 50 \\
$\xi_T$ (GHz/K) & 7.6 & $\xi_N$ (GHz m$^3$) & -3.5 $\cdot$ 10$^{-18}$ \\ 
$\Gamma_T$ (MHz) & 1 & $\Gamma_{FC}$ (GHz) & 2.2 \\ 
$g_i$/2$\pi$ (KHz) & 150 & $m_{eff}(kg)$ & 2.4$\cdot$ 10$^{-15}$\\
$\Omega_3$/2$\pi$ (MHz) & 53.15 & $\Omega_5$/2$\pi$ (MHz) & 120.05 \\ 
\end{tabular}
\end{ruledtabular}
\label{tab:example} 
\end{table}

\begin{table}[htb]
\centering
\caption{Parameters of the model after normalization.}
\begin{ruledtabular}
\begin{tabular}{lccc} 
Parameter & Value & Parameter & Value \\ \hline
$n_c$ & 2$\cdot$10$^5$ & $S$ & 10$^3$\\
$\xi_T$ & 37.92 & $\xi_N$ & -1.13 \\ 
$\Gamma_T$ & 0.124 & $\Gamma_{FC}$ & 271.6 \\ 
$g_i$ & 0.116 & Q & 500\\
$\Omega_3$ & 41.23 & $\Omega_5$ & 93.12\\ 
\end{tabular}
\end{ruledtabular}
\label{tab:example} 
\end{table}

\subsubsection*{Numerical propagation}

The normalized system of equations (\ref{eq: system2}) is solved numerically in MATLAB using standard ordinary differential equation (ODE) solving methods.  We select a temporal step of $\Delta \bar{\tau} = 4\cdot 10^{-4}$ to properly characterize the fast dynamics of free carriers and a time interval $\bar{\tau}_{span} = 10^{3}$ to ensure that the system achieves a stationary dynamical regime. The initial conditions are settled to [$\bar{N}$(0), $\bar{T}$(0), $\bar{x}_3$(0), $\bar{y}_3$(0), $\bar{x}_5$(0), $\bar{y}_5$(0)] = [0.1, 0, 0, 0, 0, 0]. \\

The input power and coupling efficiency are selected to match the one used in the experiment ($P_{in} = 3$ mW and $\eta = 0.15$). Starting from a blue-detuned position $\bar{\Delta}_0 > 0$ $(\Delta_0' < 0)$, the temporal trace of transmission (Eq. \ref{eq: transm}) is calculated in the selected time interval for each value of detuning from $-5 < \bar{\Delta}_0 < 70$ using 300 steps. The initial conditions are replaced after the first iteration with the stationary solution achieved in the previous case. \\

Figure \ref{fig: quasi}a presents the temporal traces of the computed state variables for a specific detuning where the system exhibits the quasi-periodic MML regime. Here, we observe the quasi-matching of 4 oscillations of M$_3$ for every 9 oscillations of M$_5$ within an SP quasi-cycle defined by the magnitudes $\bar{N}$ and $\bar{T}$. A longer timescale periodicity is also evident in the temporal traces associated with $\bar{x}_3$ and $\bar{T}$. We compute the temporal transmission trace (Fig. \ref{fig: quasi}b), from which a fast Fourier transform (FFT) is performed to obtain the RF spectrum for comparison with the main experiment. The temporal trace predicted by the model is also compared with experimental measurements, showing excellent agreement. This confirms that, despite its simplicity, the model captures the main physics of the system. \\

\begin{figure*}[t]
\centering    
\includegraphics[width= 0.75\linewidth]{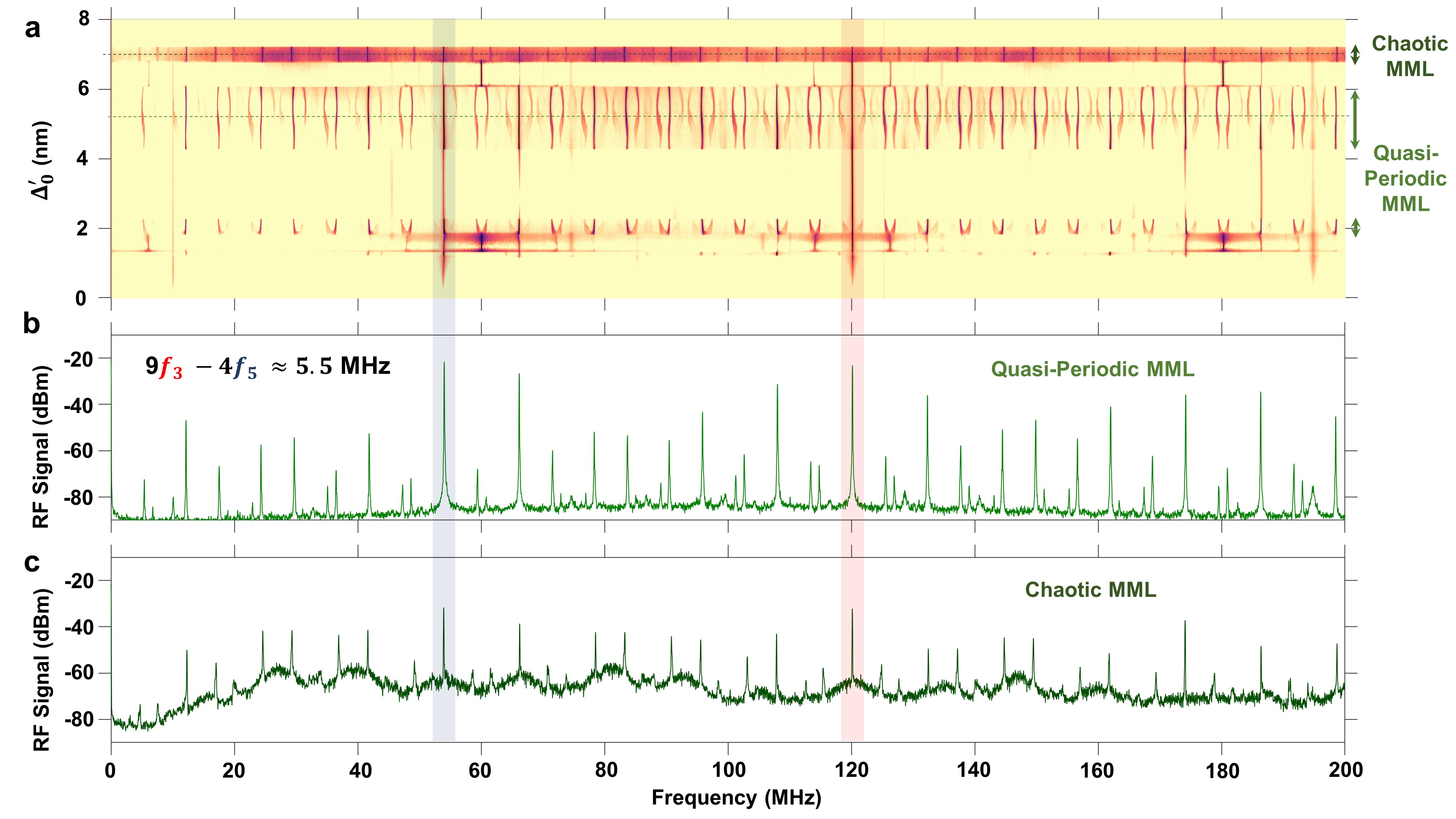}
\caption{An experiment illustrating the chaotic and quasi-periodic MML regimes in a sample where the frequencies do not satisfy the condition $9f_3 = 4f_5$. \textbf{a}  Contour plot of the RF spectra measured with a spectrum analyzer as a function of $\Delta_0'$. The quasi-periodic and chaotic MML regions are highlighted in green and dark green, respectively. \textbf{b,c} Extracted RF spectra from the quasi-periodic and chaotic MML regions shown in (a), respectively. \label{fig: quasi2}}
\end{figure*}

\subsubsection*{Bifurcation analysis}
In the previous section, we explained the methodology used for solving the system of differential equations, which consists on providing an initial condition based on the previous solution and allow the system to evolve until it reaches a stationary state. This methodology introduces two limitations. First, to reduce computational effort, we restrict numerical simulations to a timespan of hundreds of microseconds. As a result, states with small instabilities that grow over longer timescales could potentially be misclassified as stable states. Second, when providing an initial solution, the system will converge to only one stable solution, thereby losing information about the entire set of solutions that coexist within the same range of parameters. \\

To gain deeper insight into the system, we perform a bifurcation analysis of the model as a function of detuning using AUTO-07p software \cite{auto}. The results are shown in Fig. \ref{fig: bifurcations}, where the maximum values of the normalized instantaneous amplitudes of both mechanical modes are plotted for the stationary solutions and the periodic orbits emerging from two distinct Hopf bifurcations. In Fig. \ref{fig: bifurcations}a, we focus on H1, where a periodic orbit with repetition rate $f_5$ emerges. We associate this Hopf bifurcation with the onset of lasing dynamics of $M_5$ individually. During the whole detuning excursion, several period-doubling bifurcations (green dots) appear, which result in periodic orbits at $f_5/2$. Here, we identify the state R3 (see Fig. \ref{fig: fig3}a in the main text), where a solution dominated by the high amplitude of $M_5$ and a repetition rate of $f_5/2$ (red area) exits within the same detuning range observed in numerical propagation. This state corresponds to the second harmonic of the SP cycle locked to the oscillation of $M_5$. Interestingly, by analyzing the periodic orbits emerging from H2 (see Fig. \ref{fig: bifurcations}b), we observe that stable single-mode lasing of $M_3$ coexists with R3 in a certain range of detuning. This coexistence could explain the transition regimes observed in R4 or Case 1 of the main experiment (see Fig. \ref{fig: fig2}). \\

\begin{figure*}[t]
\centering    
\includegraphics[width= 0.8\linewidth]{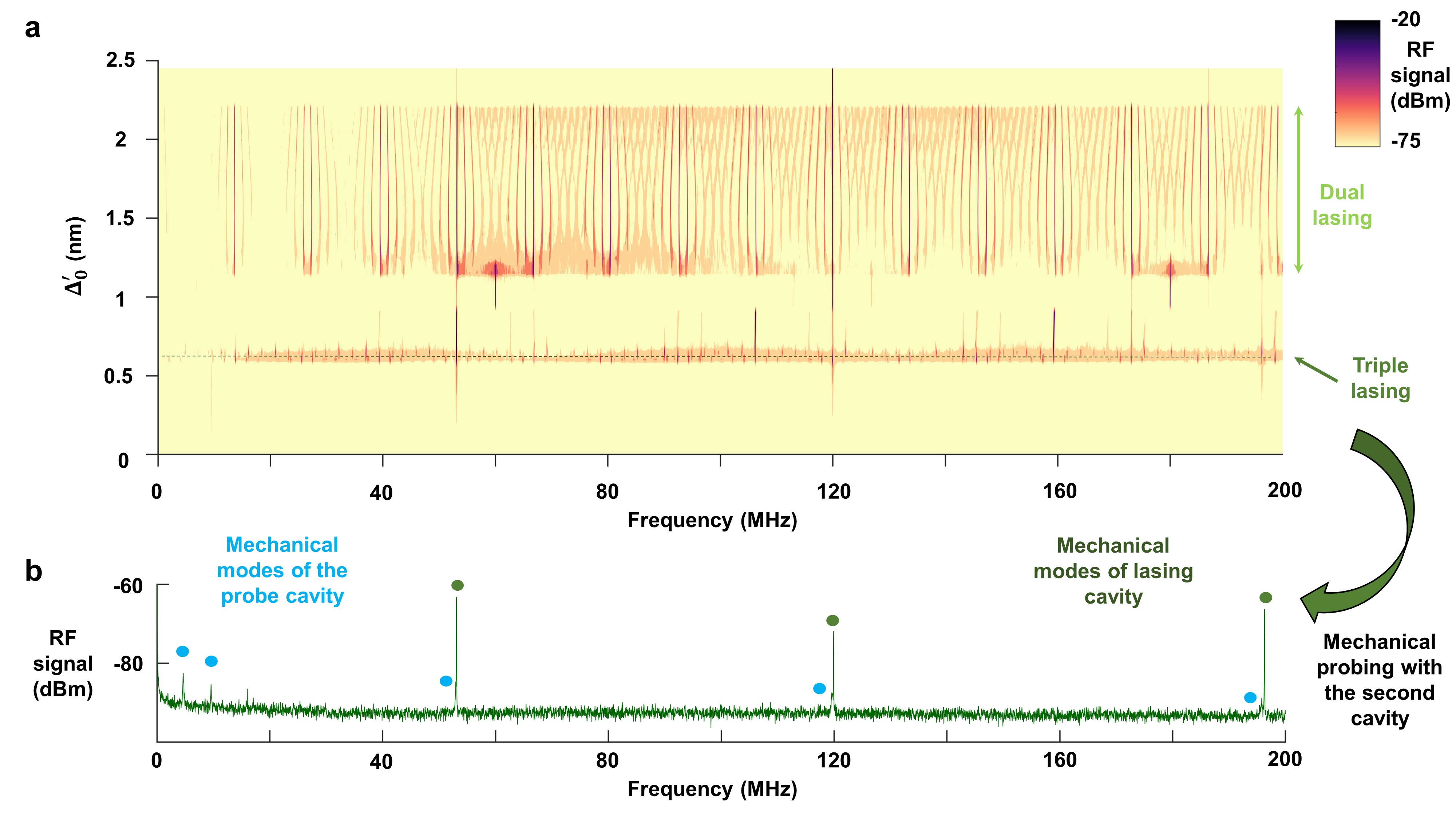}
\caption{Mechanical probing measurements of an MML state involving more than two mechanical modes. \textbf{a} Contour plot of the RF spectra measured with the spectrum analyzer as a function of $\Delta_0'$. The signal corresponds to the channel associated with the cavity exhibiting MML dynamics. \textbf{b} RF signal measured at the probing channel associated with the cavity receiving the mechanical perturbation. Only the spectrum corresponding to the three-mode lasing operation is shown. Peaks corresponding to the transduction of thermally activated mechanical modes in the probe cavity are highlighted with blue dots, while those associated with the perturbation generated by the cavity in MML operation are highlighted with green dots. \label{fig: mml}}
\end{figure*}

Beyond this region (highlighted in red), two different solutions associated with single-mode lasing of M3 and M5 coexist over a broad detuning range.  However, under certain conditions, quasi-periodic MML dynamics emerge as the stable state of the system. The Poincaré map shown in Fig. \ref{fig: fig3}b suggests that the quasi-periodicity arises from a Neimark-Sacker or torus bifurcation \cite{torus}. Despite these solutions has not been obtained using AUTO software (only observed in numerical propagation), several periodic solutions exhibiting MML dynamics with a repetition rate of $f_3/4$ are found during the numerical continuation in detuning emerging from two consecutive period doubling bifurcations (see Fig. \ref{fig: bifurcations}b). One of this solutions has been analyzed in detail in Fig. \ref{fig: fig3}c and corresponds to the locked MML state where both mechanical modes have frequencies that match the condition $4f_5 = 9f_3$.

\subsection{Flexibility of the SP cycle}
The main text shows several MML regime of quasi-periodic oscillations where the mechanical modes are close to satisfy an integer relation $f_d = 4f_5 - 9f_3 < 1$ MHz. Here, we show that SP enables the simultaneous self-sustaining of two mechanical modes in a quasi-periodic or chaotic state, even when the mismatch is much higher (around 5 MHz). \\

Figure \ref{fig: quasi2}a  shows the complete set of RF spectra as a function of $\Delta_0'$ in a contour plot representation. Similar dynamics to those presented in the main experiment are observed. Here, a complex RF signal emerges from the coherent beating between the two nonlinear, mismatched RF frequencies. Nevertheless, the peaks associated with the coherent oscillation of each mechanical mode and their frequency difference remain clearly dominant.\\

As shown in simulations (see Fig. \ref{fig: fig3}a),  prior to destabilization, the system tends to exhibit dynamics where the SP cycle displays chaotic behavior, while the mechanical modes remain self-sustained. This behavior is reflected in Fig. \ref{fig: quasi2}c, where the background RF noise increases due to the chaotic behavior of the SP cycle. Despite this, the comb emerging from the coherent beating of RF signals associated with the mechanical modes remains stable, indicating that the MML state persists, consistent with simulation results.

\subsection{Multimode lasing beyond two mechanical modes}
The dominance of an MML state over single-mode lasing operation is facilitated by the proximity of the three-antinode and five-antinode flexural modes to satisfying an integer relation between their mechanical frequencies. Here, we present a case where not just two, but three mechanical modes are simultaneously self-sustained through self-induced nonlinear intra-cavity power modulation driven by SP dynamics. \\

Figure \ref{fig: mml} presents an experiment similar to the one described in the main text (see Fig. \ref{fig: fig4}), designed to probe the mechanical signal of the nanobeam in MML operation. Throughout the detuning excursion (see Fig. \ref{fig: mml}a), multiple regimes are observed, including dual mechanical lasing in the quasi-periodic regime over a broad range. At low $\Delta_0'$ values, an even more complex regime emerges, where the seven-antinode flexural mode is also synchronously pumped, leading to high-amplitude and coherent oscillations. To confirm the simultaneous mechanical lasing of the three modes, the state is mechanically probed in the second cavity (see Fig. \ref{fig: mml}b). In this configuration, the three peaks associated with the perturbation generated by the cavity in MML operation onto the second nanobeam are clearly observed. \\ 

It is worth mentioning that this state appears only within a very narrow detuning range and is highly sensitive to room conditions, frequently hopping \cite{hopping} between different dual-lasing combinations of the three modes, indicating low stability. This behavior stems from the increased difficulty of the SP cycle to adapt and achieve synchronous pumping for all mechanical modes, given that their frequencies do not precisely satisfy an integer relation.

\end{document}